\documentclass{elsart}
\usepackage{graphicx,amssymb,amsmath,times}
\journal{NIM}
      \begin{document}
      \begin{frontmatter}
\title{ The  TACTIC  atmospheric  Cherenkov Imaging  telescope}
\author{ R. Koul},\author{A.K. Tickoo\corauthref{cor1}},\ead{aktickoo@barc.gov.in} \author{ S.K. Kaul, S.R. Kaul, N. Kumar, K.K. Yadav},
\author{ N. Bhatt, K. Venugopal, H.C. Goyal, M. Kothari, P. Chandra, R.C. Rannot},
\author{ V.K. Dhar, M.K. Koul, R.K. Kaul, S. Kotwal, K. Chanchalani, S. Thoudam},

\author{ N. Chouhan, M. Sharma, S. Bhattacharyya, S. Sahayanathan }
\corauth[cor1]{Corresponding author :}
\address{Bhabha Atomic Research Centre,\\
         Astrophysical Sciences Division.\\
         Mumbai - 400 085, India.}
\begin{abstract}
The  TACTIC  $\gamma$-ray  telescope,  equipped  with  a light collector of area 
$\sim$9.5m$^2$  and  a medium resolution imaging  camera of 349-pixels,  
has been  in operation at  Mt.Abu, India  since 2001. This paper describes  the main features  of its various subsystems and its overall  performance  with regard to (a) tracking accuracy of its 2-axes drive system, (b) spot size of the light collector, (c) back-end  signal processing electronics and topological trigger generation scheme, (d) data acquisition and control system and  (e) relative and absolute gain calibration methodology.  Using a trigger field of view of 11$\times$11 pixels ($\sim$ 3.4$^\circ$$\times$3.4$^\circ$), the telescope records a  cosmic ray event rate of $\sim$2.5 Hz at a typical zenith angle of 15$^\circ$. Monte Carlo simulation  results  are also presented in the paper  for comparing   the  expected  performance of  the  telescope  with  actual observational results. The consistent detection  of  a steady signal from  the Crab Nebula  above 
$\sim$1.2 TeV energy, at a sensitivity level of 
$\sim$5.0$\sigma$ in   $\sim$25 h,    alongwith excellent matching of its energy spectrum  
with that obtained  by  other  groups,  reassures  that the  performance  of the TACTIC  telescope is  quite  stable and reliable. Furthermore, encouraged by the detection of  strong $\gamma$-ray signals from Mrk 501 (during 1997  and 2006 observations) and Mrk 421 (during 2001 and  2005-2006 observations),  we believe  that there is  considerable scope  for the TACTIC  telescope  to monitor  similar  TeV $\gamma$-ray emission  activity from  other  active  galactic nuclei  on a long term basis.    
\end{abstract}
\end{frontmatter}
\section{Introduction}
\label{1}
Gamma-ray photons  in the very high  energy range  ( 0.1 - 50 TeV ) are expected  to come from a wide  variety of cosmic objects from, both,  within  and outside  the Milky Way Galaxy.  The candidate sources include supernova remnants [1,2], OB star associations [3], X-ray  binary systems [4,5] and active galactic nuclei [6].
Studying  this radiation in detail can yield  valuable and, quite often, unique  information  about the unusual astrophysical  environment characterizing these sources, as also on the intervening  intergalactic  space [7].   Whileas  this  promise of the cosmic TeV $\gamma$-ray probe has been appreciated for  quite a long time,  it was the landmark  development of the 'imaging'  technique and the principle of 'stereoscopic imaging',  proposed   by the Whipple [8]  and the HEGRA [9]  groups, respectively, that revolutionized the  field of ground-based very high  energy $\gamma$-ray astronomy.  In this  technique, the spatial  distribution of the photons  in the image plane (called the Cherenkov image) is recorded by using a close-packed  array of fast photomultiplier tubes 
(also called the  Imaging  Camera  with individual PMTs as its pixels).  Detailed  Monte-Carlo  simulation  studies  have shown  that the Cherenkov images   resulting  from  $\gamma$-ray  showers from  a  point source are compact and roughly elliptical  in shape  with their  major axis pointing  towards  the  source  direction in the focal plane camera [10,11]. On the contrary, Cherenkov  images resulting  from  cosmic  ray showers  are broader in size, irregular in  their shape and, are randomly oriented in the focal plane due to their  isotropic nature.   Modern  atmospheric  Cherenkov  telescopes (e.g  MAGIC [12], VERITAS [13], HESS [14] and CANGAROO [15]), utilizing  the  'imaging' technique, allow  the removal of more than 99.5 $\%$ of the cosmic-ray background,  yielding  an unprecedented  sensitivity  in the  TeV energy   range. 
\par
The TACTIC (TeV Atmospheric Cherenkov Telescope with Imaging  Camera) $\gamma$-ray telescope [16] has been in operation  at Mt. Abu ( 24.6$^\circ$ N,  72.7$^\circ$ E, 1300m  asl), India, for the last several years  to  study TeV gamma ray emission  from celestial sources.  The  telescope   uses a  tessellated  light-collector of  area  $\sim$ 9.5m$^2$ which is  capable of tracking a celestial source across the sky.  The telescope deploys a  349-pixel  imaging camera,   with a uniform pixel resolution  of $\sim$ 0.3$^\circ$ and a $\sim$ 6$^\circ$x6$^\circ$ field-of-view,   to take a fast snapshot of the atmospheric Cherenkov events produced by an incoming cosmic ray particle or a gamma ray photon with an energy above  $\sim$1TeV. 
The  photographs of the TACTIC  imaging  telescope and its  back-end  signal processing electronics are  shown in Fig.1 
\begin{figure}[h]
\centering
\includegraphics*[width=1.0\textwidth,angle=0,clip]{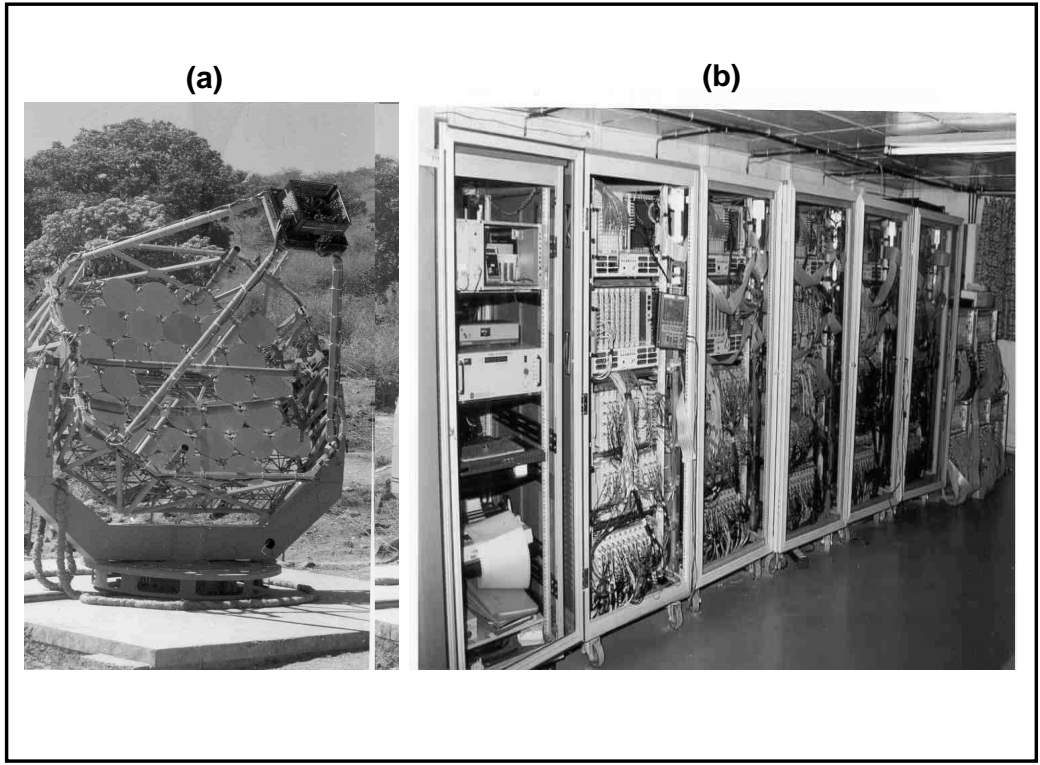}
\caption{\label{fig ---} (a) Photograph of the 349-pixel TACTIC  imaging  telescope (b) Photograph of back-end  signal processing electronics used in the telescope.}
\end{figure}
In this  paper,  we discuss the instrumentation aspects of the TACTIC imaging element.  Apart from  giving the basic description, the present paper provides  a  quantitative  performance of  its  hardware  comprising  light collector optics, drive control system,  front-end  and back-end instrumentation.  Finally,  the system performance will be evaluated through  the  
analysis  of TeV $\gamma$-ray  observations of the Crab Nebula and Mrk 421. 
\section{Observatory  Site}
\label{2}
Keeping in mind the overall scientific requirements  of  a good  astronomical site for atmospheric Cherenkov systems, a comprehensive site survey program was undertaken by us in 1993 [17] and Mt.Abu (24.63$^0$N, 72.75$^0$E, 1300m asl), a hill resort in Western Rajasthan (India) was found to be an excellent location in the country for setting up the TACTIC  telescope. 
The main reasons for choosing Mt.Abu  as the site for TACTIC  telescope  are that it offers:  (a) Maximum number of cloud free nights during a calendar year, (b) Reasonably dark site, with a negligible contribution to the background light level from artificial sources, (c)  Haze, dust and  pollution  free atmosphere. (d) Operational ease, good logistics and mild climate. 
The INSAT-1D satellite cloud cover images in the infrared (10500-12500 nm) and visible (550-700 nm) bands were studied to derive quantitative information on the percentage of clear nights per year. An analysis of the 5-year data (1986-1991), revealed that the mean percentage of clear nights at Mt. Abu is the highest, among the 7 potential sites  explored, translating to about 1170 hours per year  of effective observations  for atmospheric Cherenkov work  [17].
The monthly distribution of this observation time available at Mt. Abu is essentially uniform, except for the monsoon-affected months of June-September. Dust loads of about 20 $\mu$g m$^{-3}$, precipitable water vapour content of only a few mm during non-monsoon period and peak wind-speeds of 60 km h$^{-1}$ are some other relevant parameters of the selected site. The extinction measurements carried out at Mt.Abu  over the spectral band $\sim$ 310-560 nm lead to mean extinction coefficient of $\sim$0.47$\pm$0.10 and $\sim$0.31$\pm$0.07  for the pre-monsoon and post-monsoon period, respectively, to be compared with a value of $\sim$0.35 at Mt.Abu  altitude for a standard atmosphere. 
The mean night sky brightness levels  during the two observation spells were found to be 
$\sim(1.34 \pm 0.50)\times10^{12}$ photons m$^{-2}$s$^{-1}$sr$^{-1}$ (pre-monsoon)  and 
$\sim(1.54 \pm 0.55)\times10^{12}$ photons m$^{-2}$s$^{-1}$sr$^{-1}$ (post-monsoon).
\section{Mechanical  structure  of  the  telescope}
\label{3}
The main sub-assemblies of the TACTIC  telescope are : mirror basket, mirror fixing/adjusting frames, zenithal and azimuthal gear  assembly, encoders/motors for its two axes, camera support boom assembly and the photomultiplier tube-based imaging camera. A  three  dimensional truss type structure has been used to support the mirror frame.  The mirror basket is a 3 layer welded mildsteel  tubular  grid structure   fabricated in 3 parts  which are bolted together.  The basket is provided  with a central tie rod  which extends  into   short stub  shafts  at the two ends  which are housed in bearings. The individual  mirror facets  ( 34 in number, each with a  diameter 60 cm  and weighing around 20 kg) are supported on three levelling studs, so that desired inclination of the specific mirror with respect to the telescope axis can be achieved.   The weight of the moving part of the telescope is around 6.5 t. The complete  basket  assembly  is held  by the  fork of the telescope  which transmits  the load  through the central  roller  thrust  bearing  to the telescope foundation. The horizontal part of the fork frame is clamped on to a large gear of the azimuthal drive rotary table of the telescope.   The thrust   bearing  using  hardened rollers was specially  designed and  fabricated  in-house.   A central pin and a taper roller bearing, housed in the large gear is used to ensure proper rotation about its vertical axis. The base structure has been anchored to the ground by 4 foundation bolts. The  zenithal  motion  to the telescope is given from only one end of the basket. A common shaft assembly fixed in self aligning ball bearing having a large gear fixed on it, provides the zenithal rotation of the mirror basket. Five stage gear box providing a total speed reduction of 5189.14 has been used in the zenithal drive of the telescope.  Likewise, a four stage gear box  leading to a speed reduction of 6348.94  has been used in the azimuth drive.  A large capacity circular cable drag chain has been provided for easy and free movement of the nearly 700  signal and high voltage cables.
\par
Deflection analysis using FEM software package has also been performed at various zenith angles to arrive at the shift in the focal point position due to structural  deformations of the mirror basket frame, mirror holding attachment and support booms. The results of this study indicate that a shift of $<$ 3mm in the focal point position can be produced over the zenith angle  range of  0$^0$ to 70$^0$.
\section{Drive  control  system}
\label{4}
The   need  for using a large light collector aperture ($\sim$3.5m) and the attendant large  telescope weight 
($\sim$ 6.5 t) have led to the choice  of an altitude-azimuth (alt-azm) mounting  for the TACTIC  telescope, as against  the comparatively  simpler  equatorial  mounting.  The main advantage of the alt-azm mount is that the telescope weight  is supported  uniformly on a horizontally-placed   central thrust-bearing.  Fig.2 depicts  the block diagram of the TACTIC  drive control system  whose design is  based on the CAMAC standard.  It uses  hybrid stepper motors, sequence  generators and power amplifiers, 16-bit absolute  shaft-encoders, programmable stepper motor controllers and a GPS clock.  
\begin{figure}[h]
\centering
\includegraphics*[width=1.0\textwidth,angle=0,clip]{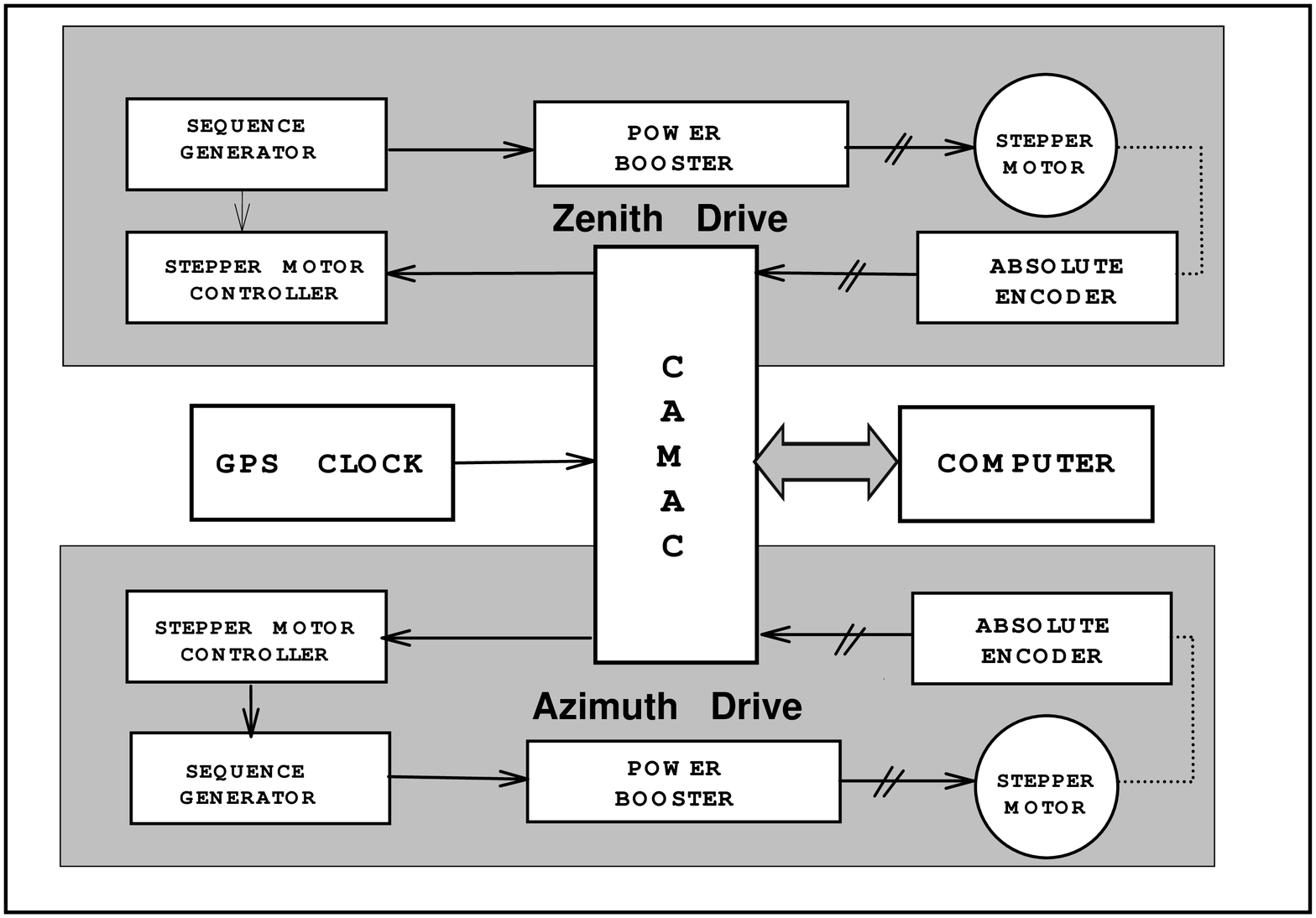}
\caption{\label{fig ---}Block diagram of the TACTIC  telescope  drive control system.}
\end{figure}
Two CAMAC-compatible, 24-bit input  registers have been used to input the telescope zenith and azimuth information to the drive system  computer to enable its control  software  to work in a closed-loop configuration  with  absolute  shaft encoders monitoring the azimuth and zenith angles  of the telescope and  the corresponding stepper motors  providing  the  movement  of its axes. 
\par
The drive torque including a reasonable safety factor of $\sim$2, required to overcome the counteracting frictional and inertial torques and for smoothly   moving the azimuthal drive  at the maximum specified  acceleration of   2 arc-min s$^{-2}$,  has been calculated  to be about 80 N cm.  This  value  has  been  estimated  on the  basis of  the expected  moment of inertia of the various moving components  of the telescope and rather conservative estimates of the coefficient of friction for the bearing used.   The TACTIC telescope  uses  two   100 N cm hybrid stepper motors (Pacific Scientific make, Model H 31NREB, NEMA  Size 34)  for driving   its  azimuthal and zenithal axes through  multistage gear-trains.  The motors   have a step size of  1.8$^0$ and are used in the 'full-step' mode.  The  speed and direction control for  each  axis of the telescope  is implemented  through an  in-house developed   CAMAC-based   Stepper Motor Controller  module  
which  provides  a   programmable  output  clock  of any desired  frequency  between 0.15 Hz and 3kHz.  
A Sequence  Generation Module,  based on the unipolar  full-step switching  pattern,   routes the input pulses to the appropriate  switches of the  chopper-based power booster.  The maximum azimuthal speed of the TACTIC telescope is 100 rad/rad (equivalent to 1500$^0$ h$^{-1}$), and the resulting blind spot has a diameter $\sim$1.2$^0$. Single turn Resolver-based  16-bit  absolute  shaft encoders (Computer Conversion Corporation make;  NEMA 12), with a resolution  of $\pm$0.33 arc-min and the worst case accuracy of $\pm$6 arc-min, are used for monitoring the orientation  of the telescope. 
The two resolvers, coupled to 16-bit resolution decoders, were calibrated on a precision indexing  table   
( resolution $\sim$0.5 arc-sec;  accuracy $\sim$5 arc-sec) and it was found that the error profiles are largely of systematic nature [18]. Accordingly, a software-based procedure  has been successfully developed for compensating for this  systematic  error-profile of the encoders.  A GPS based CAMAC-compatible  digital  clock  (Hytec Electronics Ltd. make; GPS92)  with a resolution of   $\sim$10 ns and absolute  time accuracy of   $\sim$100 ns, is used to compute  the source co-ordinates  in real time.  The new co-ordinates of the source  are calculated  after  every  second   while  tracking  a candidate source. More details regarding various  hardware  components of the  telescope  drive system  are discussed in [19].
\par 
The user friendly in-house developed  tracking  system  software provides  an independent  movement  for the zenithal  and azimuthal  axes  so that a matching  between the telescope  pointing  direction  and  the source  direction is obtained  with an accuracy  better  than  $\pm$2 arc-min. 
Once the error  goes outside the permissible bounds of $\leq$2 arc-min in case of either axis,  at a zenith angle  $\geq$7$^\circ$, a correction cycle (in the form of temporary halt or faster movement at a stepping rate of $\sim$100 Hz) is applied till the corresponding offset gets restored to within  $\leq$1.0  arc-min. 
While this  on-off type  of   correction cycle  works  perfectly for correcting the zenithal error, irrespective of the source declination,  following the  same  principle  in the azimuth  axis works  properly  for only those sources  which  have  a  minimum zenith angle of  $\>$7$^\circ$.  
For sources  which  have a  minimum zenith angle in the range 2$^\circ$ - 7$^\circ$,  
we have provided for  ramp-up correction cycle (with stepping rates upto 400 Hz), for tracking them close to their upper transit,   to avoid the problem of indefinite  chase  which  would have otherwise occurred if  the correction was performed at a stepping rate of 100 Hz.  Furthermore, since  an  azimuth error  upto  $\sim$15 arc-min  can be  easily  tolerated  at a typical zenith angle of  around  3$^\circ$   without  leading to  any serious deterioration   in the pointing  direction of the telescope,  the permissible 
azimuth error band   has been  accordingly  dynamically  widened   from   $\sim$2 arc-min  to  $\sim$15 arc-min,    depending on the  zenith angle of the source,   so that  frequent correction cycles  leading  to  a possible  'hunting' problem can be avoided.   Fig.3 gives  the representative  error profiles in the zenith, azimuth and pointing angles  of  the telescope  during  tracking  of  the Crab Nebula
(declination $\sim$22.014$^\circ$; zenith angle $\sim$2.6$^\circ$ at upper transit)  for $\sim$5 hours.
\begin{figure}[h]
\centering
\includegraphics*[width=1.0\textwidth,angle=0,clip]{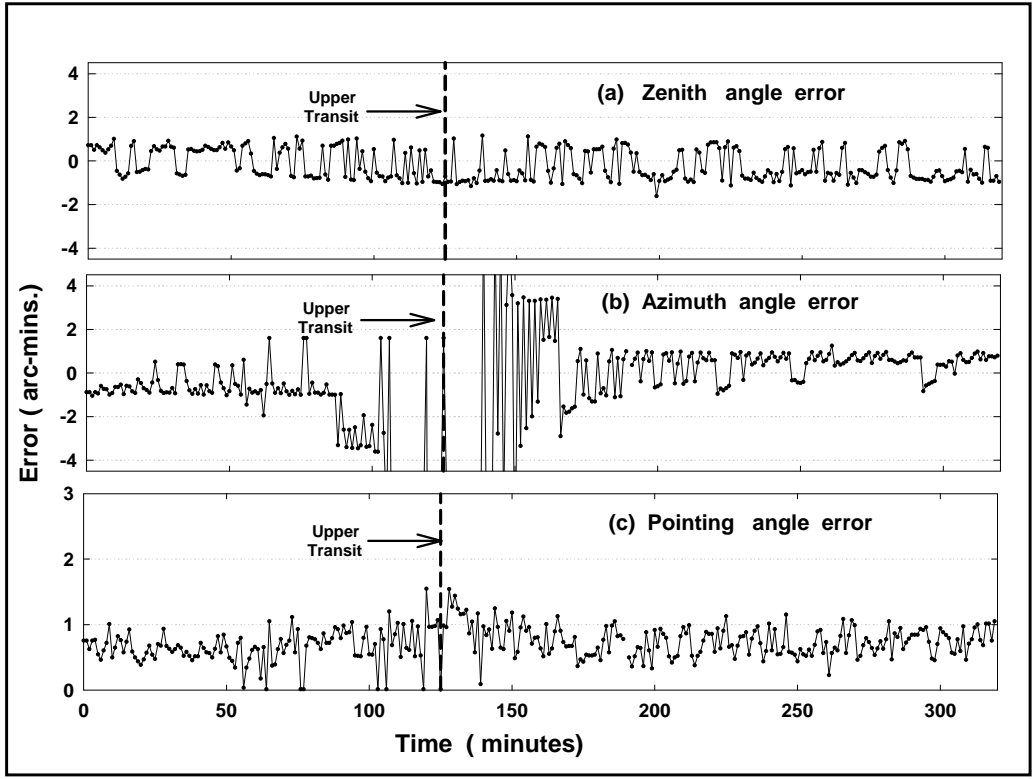}
\caption{\label{fig ---} Representative  error-profiles in the zenith (a), azimuth (b) and pointing (c)  angles  of the TACTIC telescope, obtained while tracking  the Crab Nebula (declination $\sim$22.014$^\circ$; zenith angle $\sim$ 2.6$^\circ$ at upper transit)  for $\sim$6h.  It is worth pointing out that despite encountering  large azimuth error ( i.e  around  time $\sim$ 125 minutes in the figure when the source is close to its upper transit), because of the inherent characteristics of the alt-azm drive,  the error in the pointing  direction of the telescope is still within  tolerable limits of  $\sim$2 arc-min.}
\end{figure}
Furthermore, since the TACTIC telescope  uses a single-ended drive on both axes, a correction cycle involving a fast movement along with change of direction has been deliberately avoided unless it is genuinely required for a  source, to overcome the problem of  back-lash error. 
\par 
The tracking accuracy  of the telescope  is also checked  on a regular basis  with so called 'point runs', where a reasonably  bright  star,    having  a  declination  close  to that of the  candidate $\gamma$-ray source is  tracked continuously  for about 5 hours.  The  point run calibration data  (corrected zenith and azimuth angle  of the telescope  when the star image is centered)  are  then incorporated in the telescope drive system software so that appropriate corrections can be  applied directly  in real time  while tracking  a candidate $\gamma$-ray source. 
\section{Light-collector  design  of the  TACTIC  telescope}
\label{5}
The  TACTIC light-collector   with  a collection area  of  $\sim$9.5m$^2$  uses 34   front-face aluminium-coated,  glass spherical mirrors of 60cm diameter  each  with the following characteristics (i) focal length $\sim$400cm,  (ii) surface figure
$\sim$ few $\lambda$  (iii) reflection coefficient $>$80$\%$ at a wavelength of $\sim$400nm and (iv)  thickness  20mm to 40mm.
Fig.4a  shows the projection of the mirror layout  onto  a plane transverse to the symmetry axis of the basket.
The  shorter focal length  facets are deployed close to the principal axis of the basket  while  the longer focal length facets are deployed around the periphery. The peripheral mirrors have the effect of increasing  the overall spot size as they  function in an off-axis mode.  In order to  make the design as  close to the Davies-Cotton design as possible, we  used  longer studs on the mirror frame structure to raise the pole positions of the peripheral mirrors numbered 18, 20, 26, 28, 31, 32, 33  ( shown as shaded circles  in Fig.4a). Fig.4b  shows the measured reflection coefficients of individual mirrors.  
The  reflection coefficient  measurements  of the mirrors were made  using a reflectometer (Dyn-Optics make; Model 262). The reflectometer  uses an  electronically-chopped LED-based  polychromatic  light source  with  a bell shaped spectral  distribution  extending  from  430 nm to  620 nm and having a peak emission at 550nm. Since  the reflection  coefficient  measurement involves touching the mirror surface  with a 2mm diameter bifurcated  non-scratching  fibre optic probe over  a very  small  portion of the mirror  being  tested, we have used 7 randomly selected  locations on the mirror surface  to quantify  the  mean reflection coefficient  of the mirror.   The error  bars  shown in Fig.4b  represent  standard dispersion (1$\sigma$ value) of the   7  reflection coefficient values.
\begin{figure}[h]
\centering
\includegraphics*[width=1.0\textwidth,angle=0,clip]{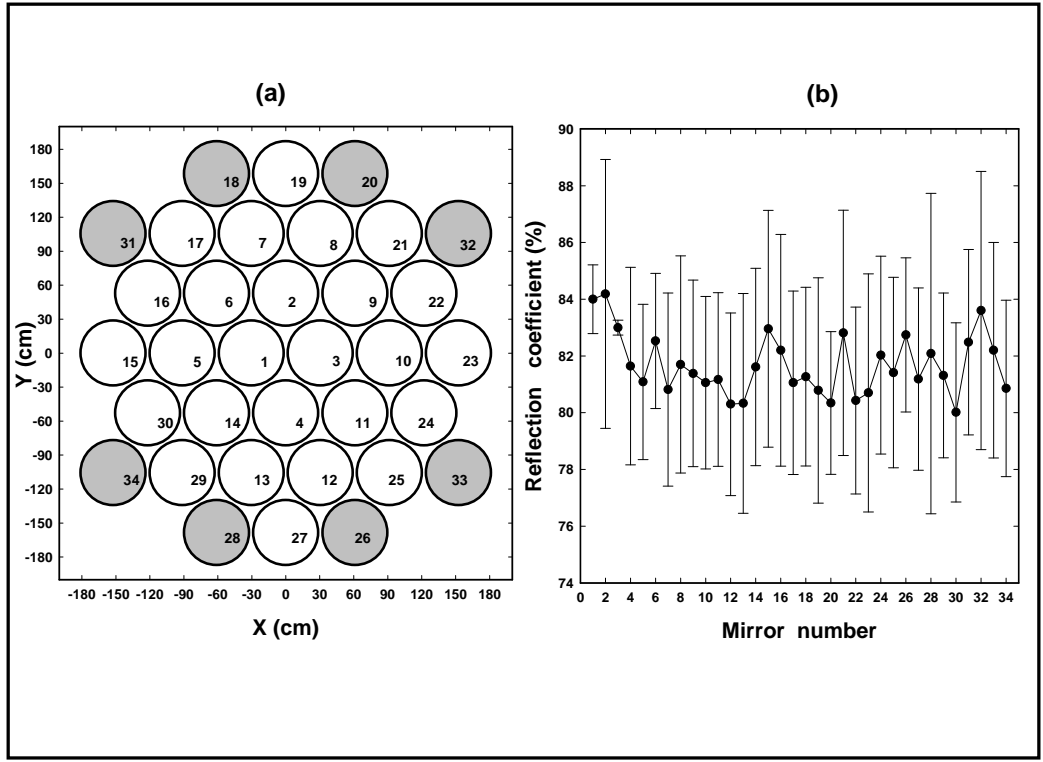}
\caption{\label{fig ---} (a) Projection of the 34 mirror facets of the tessellated light collector of the TACTIC  telescope. The shaded circles represent  the peripheral mirrors ( labelled as  18, 20, 26, 28, 31, 32, 33 and 34)  that have been  raised along the z-axis of the light collector for obtaining  a better point spread function). (b) Reflection coefficients for all the 34 mirrors used in the telescope.}
\end{figure}
\par
The alignment  of the various  mirror  facets is done by a two step process. In the first step, the orientation  angle and the focal distance of the mirror facet pole is precalculated from geometrical considerations and the orientation of the mirror to within an error of $\sim$ 1$^0$ is adjusted using a dummy facet at all the locations. The orientation of the dummy facet is set  by adjusting the 3 ball-joints which couple the triangular  mirror holding  frame to the mirror basket. In the second step, the  individual mirror facets of the telescope are further  aligned precisely  using an indigenously  developed  laser plumb-line. With  telescope pointed  in the vertical direction,  an individual  mirror  facet is installed at its pre-designated location and the laser plumb line is suspended  over it.  The mirror facet is then  slowly adjusted  such that  the reflected beam hits the  centre of the focal plane.  This fine adjustment is done by varying  the gaps at the 3 locations  below the mirror facet where it rests on the triangular holding frame.  The  above  procedure is repeated  for  a  total of 5 points on each mirror and these preselected points are  the pole of the mirror facet and four equidistant points on the periphery of the mirror.  Using  this technique for all the mirror facets, one at a time, a common focus with the minimum possible image spread was obtained  at a  focal plane distance of 386cm instead of at 400cm,  as would have been expected for a standard paraboloid or a Davies-Cotton design of the reflector. This value of focal plane distance was chosen  on the basis of the simulation results [20]. 
The   alignment of the mirror facets  is  further  confirmed  by observing a bright star image
at the focal plane. Gross misalignment in any of the facets is easily identified  as it results in  multiple 
images being seen on the focal plane.
\par
In order  to evaluate the optical quality of the  light collector experimentally,  the telescope was pointed towards the bright star  $\zeta$-Tauri and its image recorded by monitoring  the anode current of  the central pixel of the imaging  camera. The anode current versus angular offset plot  is  shown in Fig.5a. 
The  point-spread  function shown   has a  FWHM of $\sim$ 0.185$^0$ ($\equiv$12.5mm)  and  D$_{90}$  $\sim$ 0.34$^0$ ($\equiv$22.8mm).    Here, D$_{90}$  is   defined as the diameter of  circle, concentric  with the centroid of the image, within which 90$\%$  of reflected rays lie.   An image of the star Sirius recorded at the focal plane of the telescope  has also been shown in Fig.5b  and it has superimposed on it  two circles which correspond to the diameter  of the pixel and measured D$_{90}$ value calculated on the basis of Fig.5a.  The value of  D$_{90}$  $\sim$ 0.29$^0$ ($\equiv$19.3mm),  predicted on the basis of the simulation  for an incidence angle of 0$^0$,  matches  reasonably well with the measured  value mentioned above. Other details regarding the   ray-tracing simulation procedure  and comparison of the measured  point-spread function of the TACTIC  light collector with the simulated performance of ideal Davies-Cotton and paraboloid designs  are discussed in [20]. 
\begin{figure}[h]
\centering
\includegraphics*[width=1.0\textwidth,angle=0,clip]{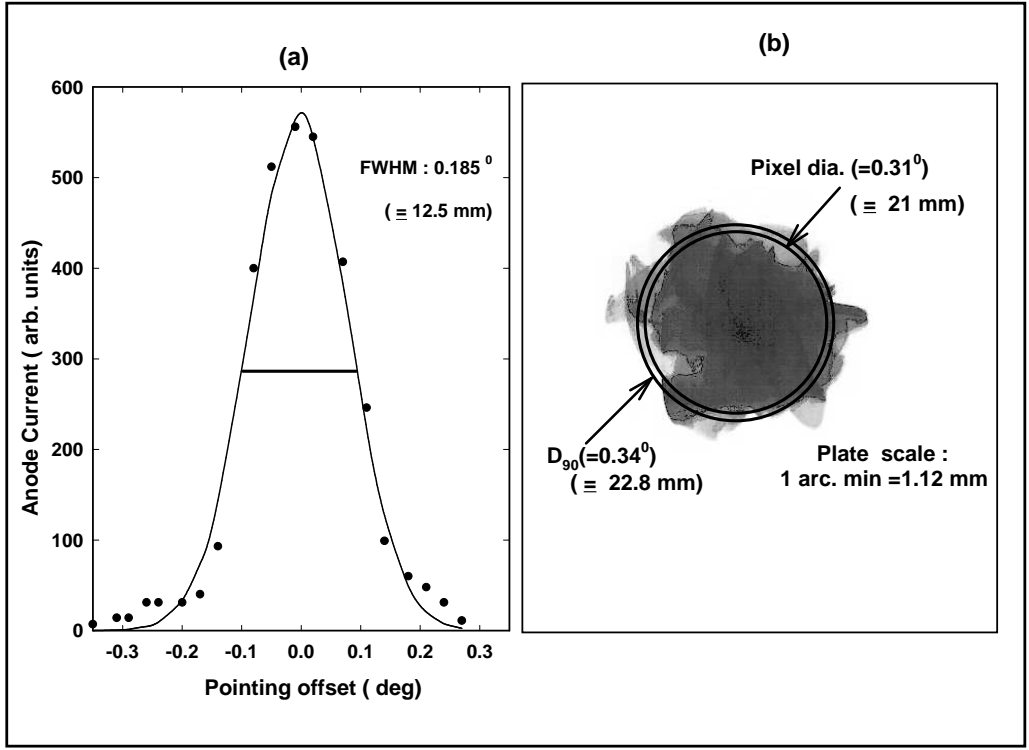}
\caption{\label{fig ---} (a) Measured point spread function of the TACTIC telescope  light collector. (b) Photograph of the image produced  by Sirius. Circles  superimposed on the image  have diameters 
$\sim$ 0.31$^0$ and  $\sim$ 0.34$^0$ and represent  the diameter of the camera pixels and  D$_{90}$, respectively.}
\end{figure}
\section{The  imaging  camera}
\label{6}
\subsection{Camera  design}
\label{6.1}
The  0.3$^0$  resolution  imaging camera  has been  designed  as a square grid  arrangement  which can be fabricated  with ease.  The camera frame is made up of  two 5mm thick  aluminium  plates  in which  19mm  diameter  holes  are drilled at a pitch of 22mm  corresponding  to the locations of the   349 pixels. The  photomultiplier  tube is held in place by a metallic  collar  fixed to its  socket  which  in turn  is held  to the rear plate by a specially designed  fastner.  Teflon  O-rings isolate the glass window of the PMT from the metallic front plate.  
The pixels  are divided  into  four sections with each section assigned  to the connector panels on one particular  side of the camera. This  arrangement  allows quick identification and replacement of malfunctioning pixels.  The central pixel of the camera which is on the  principal  axis of the light collector is used  for checking the  alignment of the mirror facets and the tracking/pointing  accuracy of the telescope.  The pixels  are numbered sequentially  clockwise  from the central pixel  which is designated  as 1.  This arrangement  has the advantage  of not disturbing the numbering subsequently as the number of operating pixels was  increased  from 81 in 1997 to 349 in the year 2001.  The  camera mounting  system  has a provision of varying its distance from the mirror basket by about  $\pm$20cm  which is very useful in optimizing  the focal plane  distance for obtaining the best possible point spread function. 
\subsection{Photomultiplier tubes and light guides}
\label{6.2}
The  imaging  camera uses 19mm diameter  photomultiplier tubes  (ETL-9083 UVB). The bialkali  photocathode has a maximum quantum efficiency of $\sim$ 27$\%$ at 340nm and the use of UV glass for the window has enhanced its sensitivity in the 280-300nm wavelength  band. The 10 stage  linear focussed  photomultiplier tube (PMT)  has a rise time of $\sim$1.8ns which is compatible with  the time profile of the Cherenkov pulse. A low current zener diode-based   voltage divider network (VDN) is used with the PMT.  This VDN design [21]   has the advantage of ensuring stable  voltages at the last two dynodes with VDN current of only about 240$\mu$A  which is a factor of 5  less than the minimum current  recommended  for a resistive  VDN.  Low thermal dissipation is an important  parameter for the stable operation of the multi-pixel  camera. The VDN uses  negative voltage and the photocathode is at a high voltage of 1000-1400V
while the anode is at the ground potential.  The  VDN of the PMT is permanently soldered to its socket  and two  RG174 coaxial cables from each VDN circuit board are terminated with coaxial connectors on the connector panels fixed to 4 sides of the camera.  The high voltage cable has a plug  type SHV  connector, while  the signal cable uses a BNC connector thereby preventing  the possibility of wrong connections. 
\par
The  Compound  Parabolic Concentrator (CPC)  shape  was chosen for the light guides  to ensure  better  light collection efficiency and reduction in the background  light falling  on the photomultipliers.  After prolonged  trials with various  materials the light guides  were  made of  SS-304.  These light guides were cut out  of rods using a numerically  controlled  lathe which  had the calculated  profile  details  stored in it.   Some of the important geometrical parameters of the CPCs  used in the TACTIC telescope camera are the following :  entry aperture   $\sim$21.0 mm; exit aperture  $\sim$15.0 mm; acceptance angle  $\sim$45.58$^0$  and height $\sim$17.6 mm. The  light collection of the CPC, which includes both the geometrical collection efficiency and the reflectivity of the surface was experimentally measured to be $\sim$65$\%$.
\section{Backend signal processing  electronics and trigger generation }
\label{7}
\subsection{Signal  processing  electronics}
\label{7.1}
The image of a typical atmospheric Cherenkov event  is registered in the form of varying amplitude pulses of $\sim$ 3mV  - 60 mV produced by a group of  5-20 pixels of the camera.  These voltage pulses are brought to the control room,  using 55m long high quality RG 58 coaxial cables.  In-house developed  fast NIM  Hex amplifier modules with a user selectable gain range of 2-50 and amplitude discriminator modules of 50 -500mV range are used for amplification and threshold  selection of the PMT signals. 
While the multichannel fast NIM-based  amplifier and fixed threshold discriminator modules
have front panel adjustments for gain and threshold respectively, the charge content and the scaler rates are directly read off the CAMAC bus. 
The complete back end instrumentation  (Fig.6) based on inhouse developed medium channel density modules is housed in  seven 19 inch racks of 36 U height (Fig. 1b).  
\begin{figure}[h]
\centering
\includegraphics*[width=1.0\textwidth,angle=0,clip]{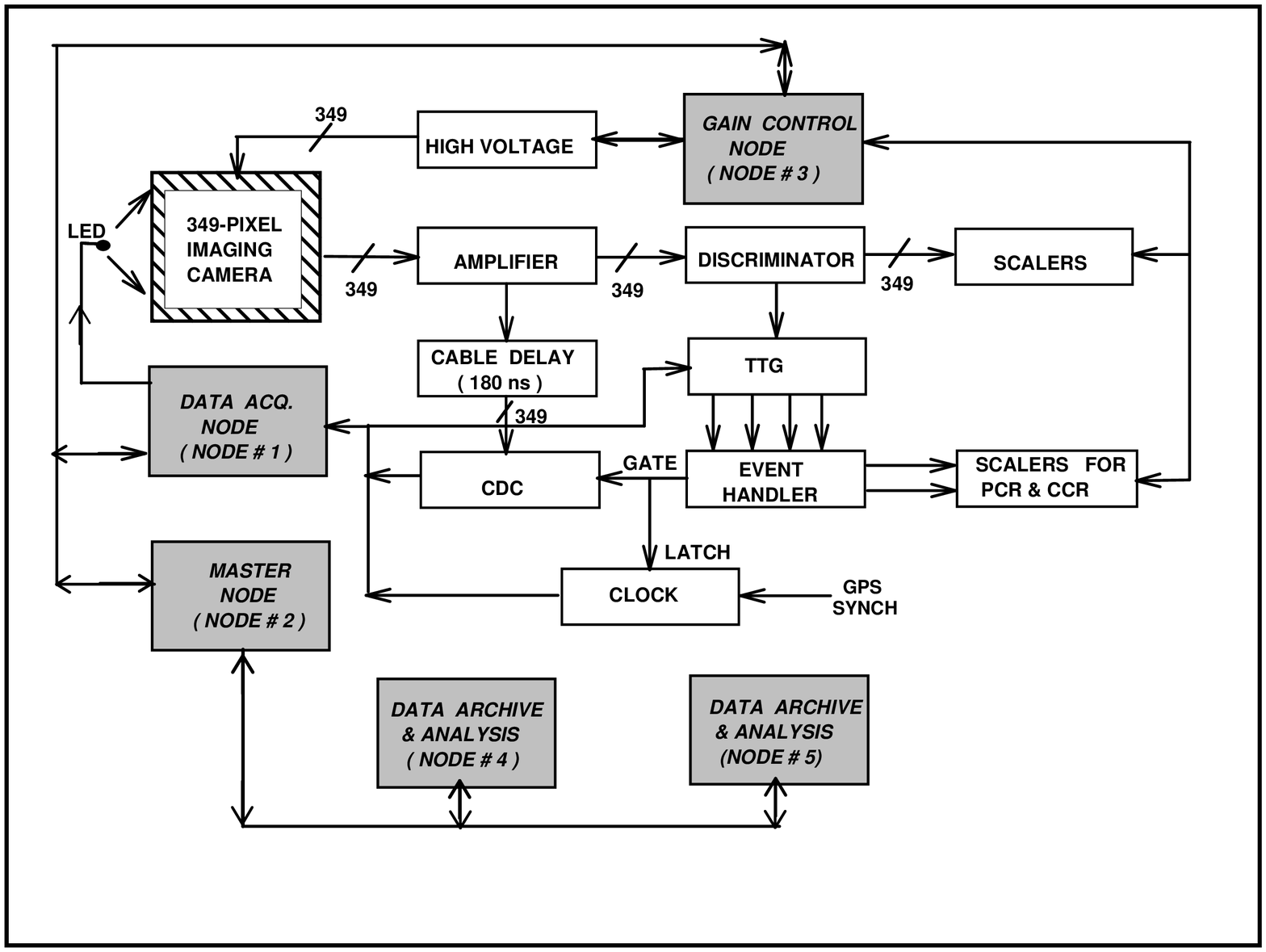}
\caption{\label{fig ---} Block diagram of the back-end  signal processing  electronics  used in the TACTIC  imaging element; TTG - TACTIC  Trigger generator; PCR-Prompt Coincidence Rate ; CCR- Chance Coincidence Rate.}
\end{figure}
One of the two outputs of a discriminator channel is used for monitoring the single channel rate with a CAMAC scaler while the other output  is connected to the  trigger generator  for  trigger generation.  The  outputs   from  each of the four independently operating   trigger generator modules  are then collated  in an Event Handler which generates  the 22ns  duration gate pulse  and interrupts  the data acquisition  for  reading  the charge ADC data  from all the 349 pixels.  The  final trigger  pulse  is also  used  for   latching  the   GPS-based   clock.  The scalers and charge-to-digital converters (CDC) for the 349 channels use 5 CAMAC crates each.  Each of these crates is controlled using an in-house developed multicrate  CAMAC controller and  five such controllers are daisy chained and connected to a data acquisition PC.  A  similar  strategy  following a custom built standard has been used for the computer-programmable high voltage units. 
\subsection{Trigger generation }
\label{7.1}
The imaging  camera  uses  a programmable  topological trigger  [22] which  can pick up  events  with  a variety of
trigger configurations. As  the trigger  scheme is not  hard wired,  a number of coincidence  trigger options 
( e.g Nearest Neighbour Pairs,  Nearest Neighbour  Non-collinear Triplets and Nearest Neighbour  Non-Collinear  Quadruplets)  can be generated under software control.  The  layout of the 349-pixel TACTIC imaging camera, which can  use  a maximum  of 240  inner pixels for trigger generation is  depicted in Fig.7a.
\begin{figure}[h]
\centering
\includegraphics*[width=1.0\textwidth,angle=0,clip]{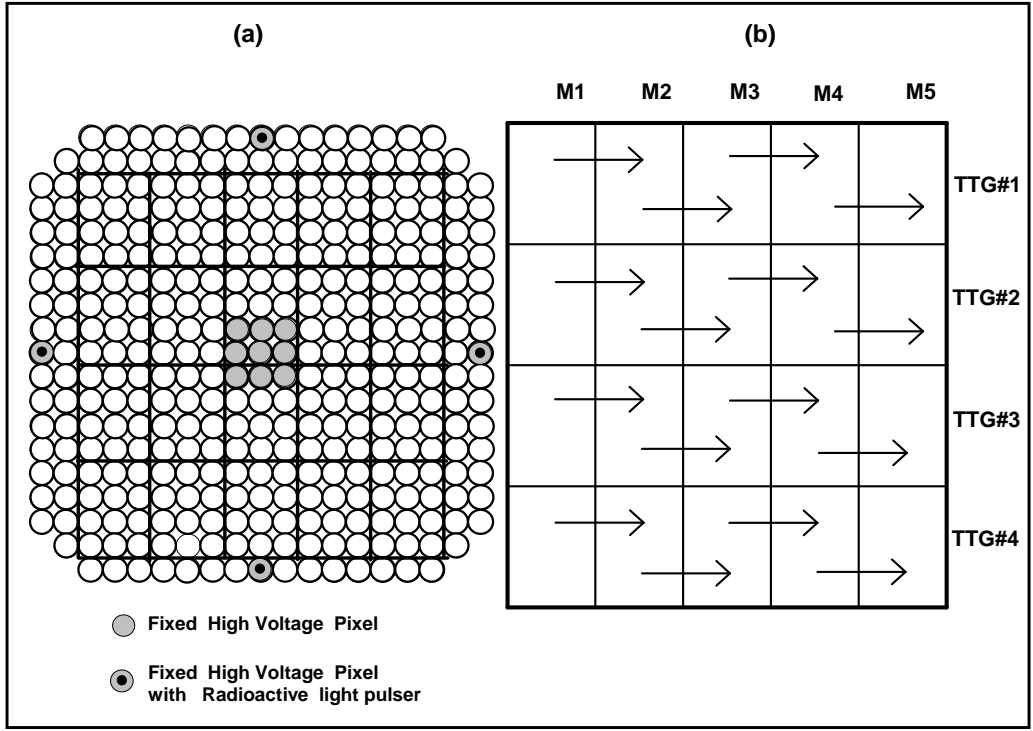}
\caption{\label{fig ---} (a) Layout of the  349-pixel imaging camera of the TACTIC  telescope. Shaded pixels in the centre and the same  filled  with circles at the periphery of the camera  represent  PMTs for which single channel rate is not stabilized. (b) Pictorial representation of the horizontal cascading followed  within each  trigger generator module. Each trigger generator module  can handle 60 channels  by using 5  memory ICs (M1-M5)  of   type TC55B417.} 
\end{figure}
The  trigger  criteria  have  been implemented  by  dividing   these inner  240 pixels 
into 20 groups  of 12 (3 $\times$4) pixels.  A section of  5 such groups  is connected to a TACTIC Trigger Generator (TTG) module and a total of 4 TTGs are required  for a maximum  of   15$\times$16 matrix of trigger pixels.   
The trigger  scheme   has been  designed  around  16k x 4 bit  fast static RAM 
(Toshiba make TC55B417; access time of $<$ 8ns).   
Each  of the 4 TTG  modules  uses 5  memory ICs (indicated by  M1-M5  in Fig.7b) and  has  horizontal cascading built into it.  A pictorial representation of the horizontal cascading  followed within each trigger generator module  is  shown in  Fig.7b.   Vertical cascading has not been provided in the trigger generator due to
non-availability  of more than 3 outputs from the discriminator. The loss of events   as a result   of  the absence of  vertical  cascading  has been estimated  to be  about 12$\%$  through Monte-Carlo simulations and actual experimentation [23]. 
The memory address lines are connected to the CAMAC address lines and the front panel receptacles ( for connection to the  discriminator outputs) through two sets of tri-state buffers. The TTG operation starts with writing of the data, as per a user defined topology,  from a disk file into each of its memories under CAMAC control.  
Once programmed,  the TTG outputs follow the event topology  as  described earlier. Apart from generating  the  prompt trigger,  the trigger generator has a provision for producing a chance coincidence output based on  $^{12}$$C_{2}$  combinations from various groups of closely spaced 12 channels.   This chance coincidence output is used as a system monitor for evaluating its overall functioning during an observation run.  Monitoring of the chance coincidence rate has also helped  in keeping a close check on the operation of the telescope and the quality of the data collected by it. 
Other details  regarding the  design, implementation and performance evaluation  of the programmable topological trigger generator  for the 349-pixel imaging camera of the  TACTIC  telescope  are discussed in [23]. 
\section{Data acquisition and control system}
\label{8}
\subsection{System architecture}
\label{8.1}
The data acquisition and control system of the telescope has been designed around a network of  PCs running the QNX (version 4.25 [24]) real-time operating system.   The software is designed for the real time acquisition of event and calibration data and on-line display of telescope status in terms of prompt and chance coincidence rates and the functional status of each of the 349 pixels of the camera.   The QNX operating system was chosen for its multitasking, priority-driven scheduling and fast context switching capabilities.  In addition,  the operating system also provides a  powerful set of interprocess communication capabilities via messages, proxies  and signals.   The data  acquisition and control  of the TACTIC is handled by a network of three  personal computers as shown in Fig.8. 
\begin{figure}[h]
\centering
\includegraphics*[width=1.0\textwidth,angle=0,clip]{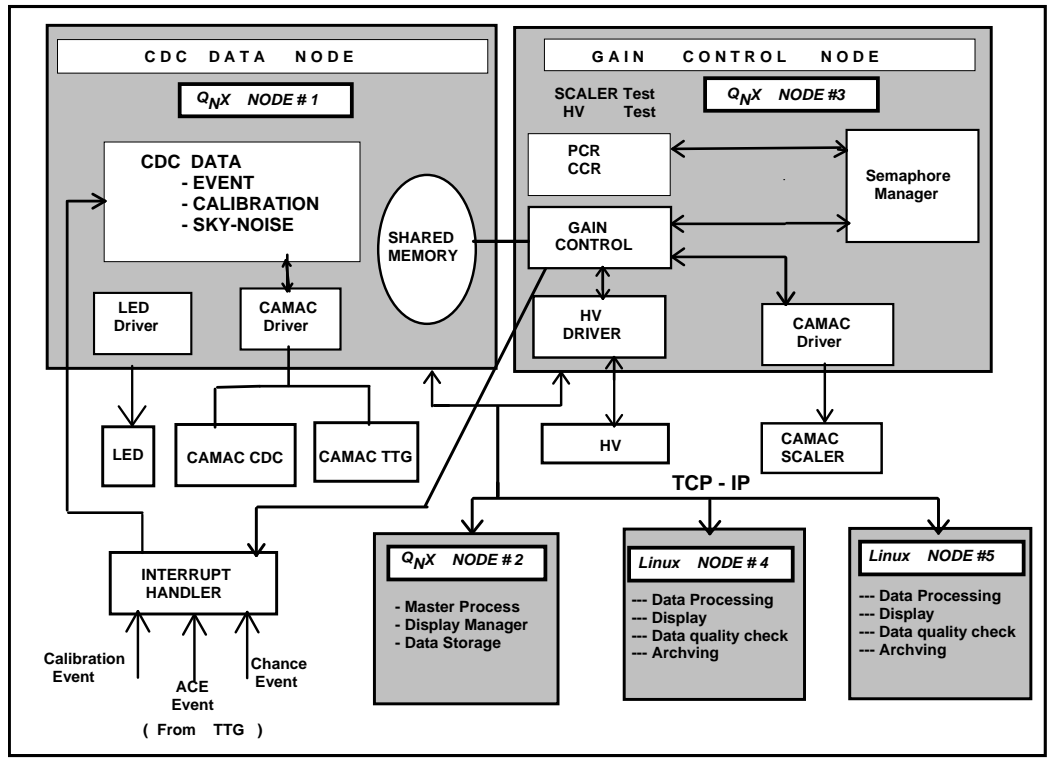}
\caption{\label{fig ---} Block diagram  of the multi-node  PC-based  data acquisition and control system.}
\end{figure}
While  one PC  is used to monitor the scaler rates and control the high voltage (HV)  to the photomultipliers,  the other PC handles the acquisition of the event and calibration data and the programming of the TTGs.  These two front-end PCs,  referred to as the rate stabilization node and the data acquisition node respectively,  along with a master node form the multinode Data Acquisition and Control network of the TACTIC Imaging telescope. All executable routines stored on the master node are spawned on to the other two front-end nodes as and when required.
The same network is extended to two more LINUX-based PCs which are used for on-line data analysis and archiving.
An event handler  module controls the whole process of data acquisition  and also provides  the link between the  TACTIC hardware and the application software.  The event handler accepts the atmospheric Cherenkov events, calibration and chance trigger outputs from various TTG modules and interrupts the front end data acquisition node.  
The system acquires the 349 channel CDC data for the trigger selected atmospheric Cherenkov events, relative calibration flashes generated by the calibration LED and sky pedestal events, in addition to CDC data for the 4 absolute calibration channels.  The high voltage  and scaler data are also logged continuously,  though at a much lower frequency.   
At event occurrence the event handler  also generates a TTL output for latching the system clock and a 20 ns wide NIM pulse for gating the CDC modules. 
The dead time of the system has been experimentally measured to be $\sim$2.5ms by collecting the relative calibration data  along with absolute time information at a trigger rate of $\sim$400Hz. Other details regarding  hardware and software features of the  data acquisition and control system of the telescope are discussed in [25].
\par
A platform-independent web-based system using the concept of `virtual instrumentation console' is being developed to provide interactive control of the telescope from Mumbai  which is about 800 km away from the Mt. Abu observatory. The remote control system will provide location independent access to the data acquisition, control and analysis resources at the observatory using a dedicated 64/128 Kbps ANUNET (Satellite based network) link. The system has two layers with the bottom layer having QNX based telescope data acquisition system linked to WinNT based web server using TCP/IP socket programming. The top layer has java client-server application using servlet communication to provide a rich user interface through standard internet browser. A remote daemon running in the background of the master node will accept connection requests from the web server and once the connection is established it will transmit current status information of the observation run to the remote client. X.509 certification will be used for server and client authentication. The web-server machine will use two network cards and a proxy server installation for internal network security from the outside world. In order to ensure effective utilization of the satellite network bandwidth the data and commands entered by the user on the virtual console will be transmitted to the remote site using very small data volume. Apart from the telescope control the system can also be used for audio-video interactions among the scientists at the two locations. 
\subsection{ Stabilization of single channel rates}
\label{8.2}
A cost-effective  method  for operating  the  imaging camera of the TACTIC $\gamma$-ray telescope at stable single channel rates (SCR) and safe anode current values  is being  used  despite  variations in the light of the night sky experienced by the individual  pixels  from time to time [26].  The camera  operates  13 PMTs ( 9 in the central region and 4 in the periphery of the camera) with fixed high voltages and the remaining  336 pixels at different high voltages to ensure  their operation within a pre-determined  Single Channel Rate (SCR)  range.  
The  purpose  behind using the central  9 pixels  of the camera at fixed high voltages is to facilitate the gain normalization (flat fielding) of the remaining 336 camera pixels, so that the  event sizes ( $\equiv$ sum of CDC counts in the clean, flat-fielded image) recorded during a nights observations can be directly compared to one another.  
Operation of the pixels in a narrow SCR band  has the advantage of ensuring  a stable chance coincidence rate  which can be used  as a system diagnostic parameter.  An  elaborate  algorithm [26] has been developed to monitor the SCR rates of all pixels using the CAMAC front ends and   ensure their operation within a narrow range despite changes  in  the background light level incident on them due to changes in the  sky brightness and star-field  rotation.  The algorithm  also ensures that all the pixels of  the camera operate within  safe anode current ranges. The feedback loop of the algorithm  changes the high voltage to the various pixels using  multichannel  high voltage unit  which has a resolution of 1V.
The decision of operating  a pixel under enhanced light levels is solely based on the comparison of the SCR and applied high voltage with reference data generated under controlled light level conditions.
A detailed description of the single channel rate stabilization  scheme  can be found in [26].
\section{ Relative  and absolute gain calibration  scheme}
\label{9}
The PMT calibration scheme employed for TACTIC has two parts, viz., relative gain calibration  and absolute gain calibration. In the relative gain calibration scheme, we use a high intensity  blue LED ( Nichia Japan make  SPB 500) at a distance of $\sim$2m from the camera to determine the relative gain of the imaging camera pixels. The LED has been  provided with a light-diffusing medium  in front of it to ensure the uniformity of its  photon field  within $\sim$$\pm$ 6$\%$.  The mean light intensity  from the pulsed LED recorded by each pixel, in response to 2000 light flashes is  subsequently used for off-line relative gain calibration of the imaging camera. 
\par
The absolute gain calibration system of the camera  involves  monitoring  the  absolute  gain  of  a  set of  4 gain calibrated  pixels (shown by  filled circles at the periphery of the camera in  Fig.7a). Since measurement of  the absolute gains of these PMTs  by determining  their  single  photoelectron peaks  a number of times  during  an observation run is rather time consuming, we have instead  used a relatively simpler method  of  measuring the light pulser yield of a calibrated  source  for  the  in-situ  determination  of the absolute gain  of  these calibration channels.   
The calibrated  light  sources used are the    Am$^{241}$-based  light pulsers (Scionix Holland BV make; dimensions of YAP:Ce pulser units - 4 mm$\times$1 mm)    which  produce    fast  optical  flashes at an average rate of $\sim$20 Hz  with maximum emission at a wavelength of $\sim$370nm.  After taking several measurements of single photoelectron peak and the radioactive  light  pulser (RLP)  yield    under dark room conditions to  validate the  reproducibility of the  measurements and  for preparing the reference data base,  
the PMTs of these calibration channels  are  mounted  permanently with  radioactive light pulsers.  
A collimator is  also used during the mounting  of a light pulser on a particular  tube so that the number of photoelectrons per pulse is  $\sim$500 pe.
These calibration pixels are  operated at fixed high voltage values and any changes in the RLP  yield  measured  during actual  observations, are  then attributed to actual gain change of the pixels.  Since  the  4 calibration pixels are also partially exposed  to the light flashes from the LED during the relative calibration run,  it becomes possible to determine  the  gain  of  all  the pixels of the camera. A representative example of single photoelectron  response  of  one of the PMT's  is shown in Fig.9a.
The mean amplitude of the single photoelectron peak (indicated  by A$_{SPE}$ in  
Fig.9a)  is then determined by fitting a Gaussian distribution function to the differential rate curve. 
Fig.9b gives the pulse height distribution of the  light  flashes  obtained  with one of the radioactive light pulsers.
\begin{figure}[h]
\centering
\includegraphics*[width=1.0\textwidth,angle=0,clip]{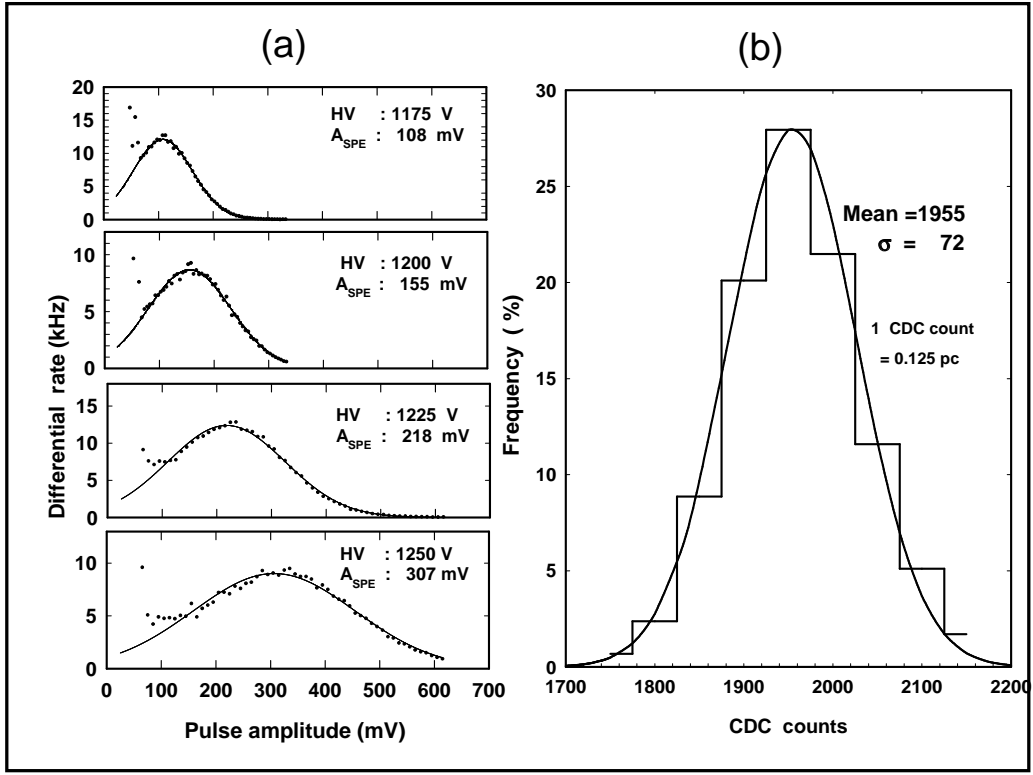}
\caption{\label{fig ---} (a) An example of single photoelectron peak  obtained for one of the calibration pixels at different values of HV. (b) Representative example of the pulse height distribution obtained with a   Am$^{241}$-based light pulser.}
\end{figure}
The underlying principle   for converting  the  charge content of  an  uncalibrated  pixel from  CDC counts to  photoelectrons,  uses the fact  that the calibration pixels are also exposed to the light flashes from the LED during the relative calibration run and hence it becomes possible to obtain the conversion factors for all the remaining  345 pixels of the camera [27].
The conversion for image size in CDC counts to number of photoelectrons  has also been  performed  independently  by using the excess noise factor method. The analysis of relative calibration data yields  a value of  1pe $\cong$  (6.5$\pm$1.2) CDC for this  conversion factor when an average value of $\sim$1.7 is used for excess noise factor of the photomultiplier tubes. Additional work   to  determine the excess  noise factor values  more precisely  for  TACTIC photomultiplier tubes is  still underway. 
\section{Monte Carlo simulations and comparison with real data}
\label{10}
\subsection{ The simulation chain}
\label{10.1}
Due to the non-availability  of a calibrated  beam of very high energy $\gamma$-ray photons, detailed Monte Carlo simulations  offer the only way to bench mark the design and performance of an atmospheric Cherenkov imaging telescope.  Measurements of absolute $\gamma$-ray flux and energy spectra of established $\gamma$-ray sources, as well as determination of upper limits on $\gamma$-ray emission from quiet objects also rely heavily on Monte Carlo predictions. 
We have  used  the CORSIKA (version 5.6211) air shower simulation code  [28] for  predicting  and optimizing  the performance of the TACTIC imaging  telescope. 
The  complete execution of the Monte Carlo simulations for TACTIC  telescope was  subdivided into two steps. The first part comprised  generating the air showers induced by different primaries and recording the relevant raw Cherenkov data (data base generation). Folding in the light collector characteristics and PMT detector response was performed in the second part. 
The simulated data-base for $\gamma$-ray showers used  about 34000 showers in the energy range 0.2-20 TeV  with an impact parameter of 5-250m. These showers have been generated at 5 different zenith angles ($\theta$= 5$^\circ$, 15$^\circ$, 25$^\circ$, 35$^\circ$ and  45$^\circ$). A data-base of  about 39000 proton initiated  showers in the energy range 0.4-40 TeV, were  used  for studying the  gamma/hadron separation capability of the telescope.  
The  incidence  angle  of the proton showers  was simulated  by  randomizing  the shower directions in a field of view of 6$^\circ$x6$^\circ$  around  the pointing direction of the telescope.  
The Cherenkov   photons are ray-traced to the detector focal plane and the number of photoelectrons likely to be registered in a PMT pixel are inferred after folding in the relevant optical characteristics of the mirrors, the metallic compound-paraboloid light concentrator at the entrance window of the pixels and the photocathode spectral response. The Cherenkov  photon data-base, consisting  of number of photoelectrons registered by each pixel,  is  then subjected to noise injection, trigger condition check and image cleaning. 
The resulting data-bases, consisting of pe distribution in the imaging camera  at various core distances and zenith angles  are   then used  for estimating (a) trigger efficiency (b) effective detection area (c) optimum ranges of Cherenkov image parameters for discriminating between $\gamma$-ray and cosmic ray events (d) differential count rate for $\gamma$-ray and cosmic ray events and (e) effective threshold energy of the telescope  for $\gamma$-ray and cosmic ray proton events.
The clean Cherenkov images were characterized by calculating their standard image parameters like LENGTH, WIDTH, DISTANCE, $\alpha$, SIZE and   FRAC2 [10,29]. The standard Dynamic Supercuts [30] procedure was  used to separate $\gamma$-ray  like images from the  background cosmic rays. 
The effective collection area of the telescope, for $\gamma$-ray and proton events, at  two representative zenith angles  values  of 15$^\circ$ and  35$^\circ$  is  shown  in  Fig.10a and 10b, respectively.  
\begin{figure}[h]\centering
\includegraphics*[width=1.0\textwidth,angle=0,clip]{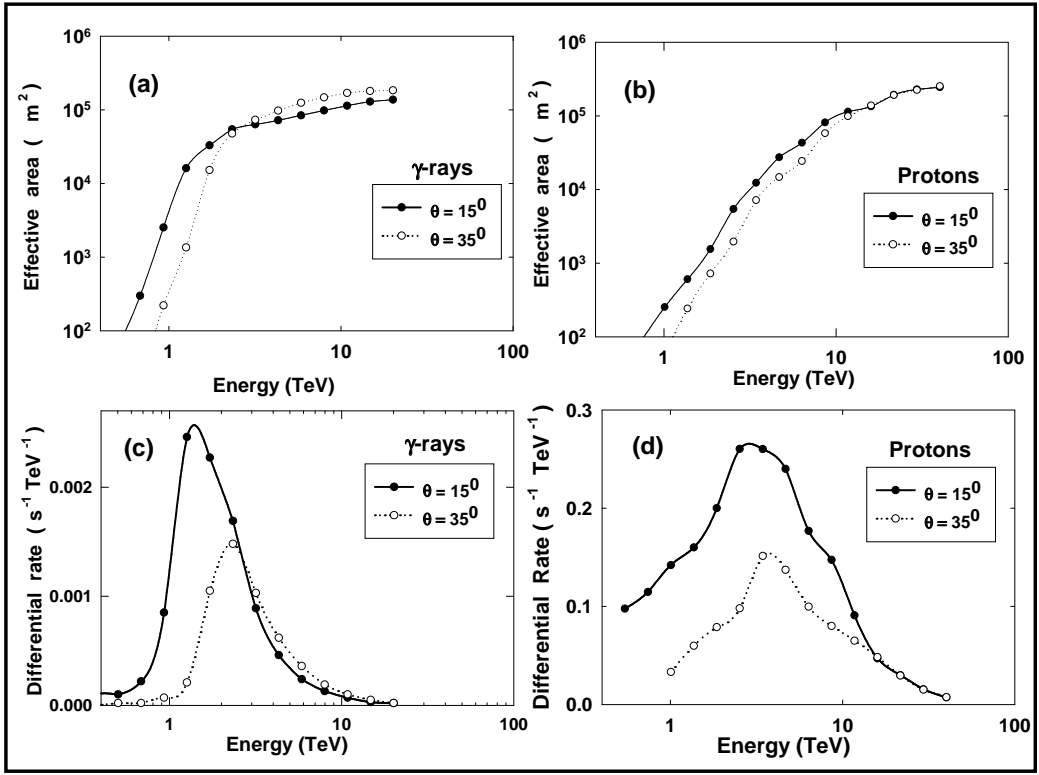}
\caption{\label{fig ---}  Effective collection are of the telescope for $\gamma$-rays(a)  and protons (b)  as a function of the primary energy at zenith angles of 15$^\circ$ and  35$^\circ$.  Differential trigger  rates for  $\gamma$-rays (c) and proton events (d)  as a function of the primary energy.} 
\end{figure}
These results  were obtained  by using the nearest neighbour  topological-trigger  with  11$\times$11  trigger field  and   a single pixel threshold  of  
$\geq$25 pe.
Fig.10c and 10d  show  the corresponding  differential  event  rates as a function  of the primary energy  for $\gamma$-ray and proton events, respectively.  Defined as the energy where the differential  rate peaks and assuming a  Crab Nebula type of spectrum with a differential exponent of  $\sim$ -2.62 [31], it is evident from Fig.10c  that  the $\gamma$-ray  trigger threshold energy of the telescope is  $\sim$ 1.2 TeV. The  corresponding trigger threshold  energy  of the telescope for protons turns out to be  to $\sim$ 2.5 TeV  (Fig.10d).  
\subsection{ Comparison  with  real data}
\label{10.2}
The  agreement between the predictions from Monte Carlo simulations and the actual performance of the telescope was first checked by comparing the observed trigger rate of the telescope with the predicted value. 
The expected prompt coincidence rate at a zenith angle  of 15$^\circ$  turns out to be $\sim$ 2.5 Hz for the  nearest neighbour pair  trigger mode.  This value  has been  obtained  on the basis of integrating  the differential  rate  curve for protons (Fig.10d).  Reasonably good  matching of this with the experimentally  observed value of $\sim$ 2-3 Hz suggests that the response of the telescope is very close to that predicted by simulations.   
\par
The  agreement between the  expected and  actual  performance  of the telescope  was next checked  by  comparing the    expected  and observed  image parameter distributions. Fig.11 shows the distributions of the image parameters  LENGTH, WIDTH, DISTANCE and  $\alpha$  [10,29] for simulated protons and for the actual Cherenkov images  recorded by the telescope.  The simulated  distributions of these image parameters for $\gamma$-rays   have  also been shown in the figure for comparison. 
\begin{figure}[h]
\centering
\includegraphics*[width=1.0\textwidth,angle=0,clip]{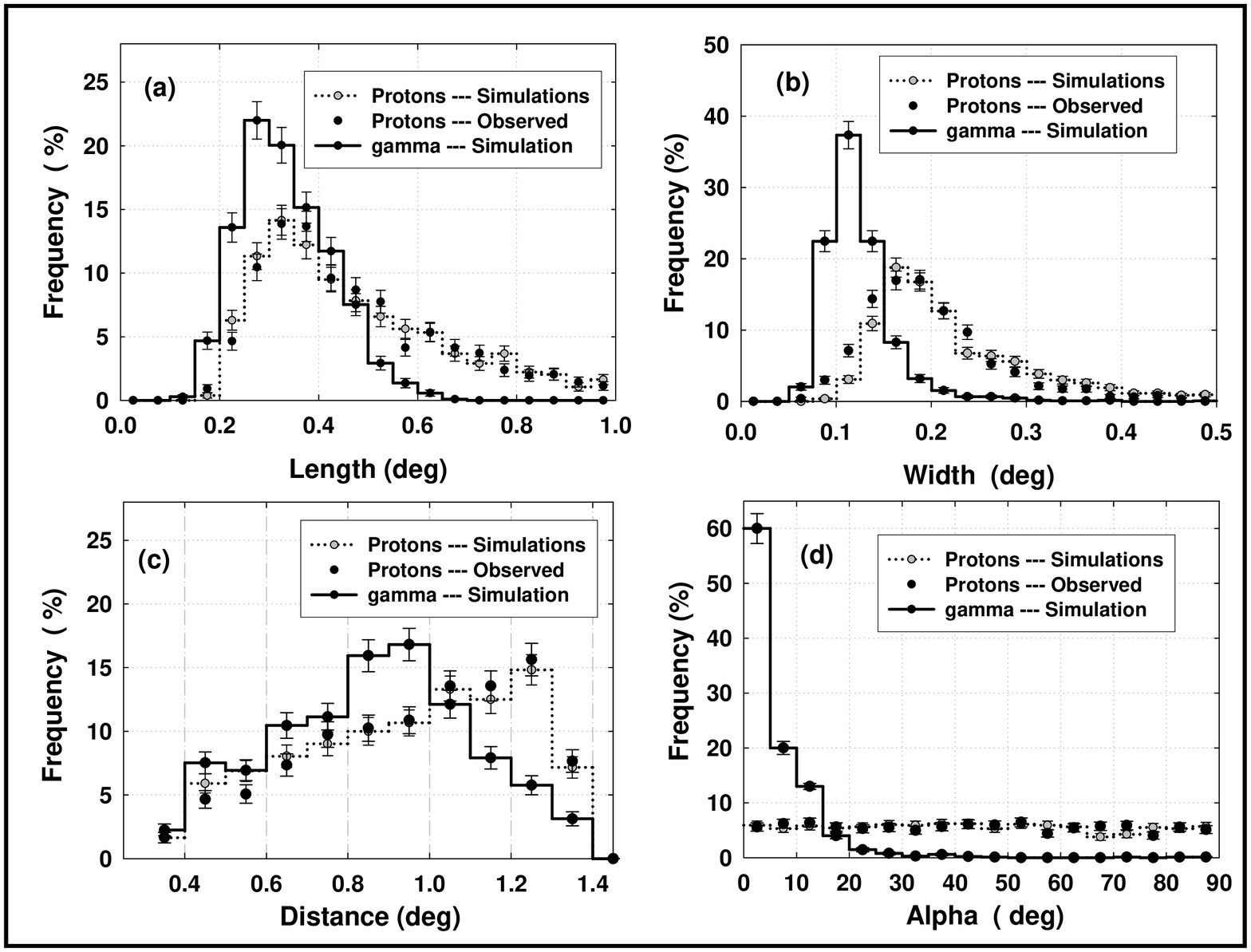}
\caption{\label{fig ---} Comparison of image parameter distributions  ( (a) LENGTH, (b) WIDTH, (c) DISTANCE and 
(d) ALPHA )  from real and the Monte Carlo simulated data for proton events. The simulated  image parameter distribution of  $\gamma$-rays  has also been shown in the figure for comparison.} 
\end{figure} 
The observed image parameter distributions are found to closely match the distributions  obtained from simulations for proton-initiated showers, testifying to the fact that  the event triggers are dominated by background cosmic rays. 
\par
A novel  ANN (Artificial Neural Network)-based  energy estimation procedure  for determining the energy spectrum of a candidate $\gamma$-ray source has also been developed. The procedure followed  by us uses a 3:30:1 (i.e 3 nodes in the input layer, 30 nodes in hidden layer and 1 node in the output layer) configuration of the ANN with resilient back propagation  training algorithm  to estimate the energy of a $\gamma$-ray like event on the basis of its image SIZE, DISTANCE and zenith angle.  The new ANN-based energy reconstruction method developed by us, apart from yielding a  $\sigma$(lnE) of $\sim$28.4$\%$,  has the added advantage that  the  procedure  allows data collection over  a much wider zenith angle range as against a  coverage of upto 35$^\circ$ in case the zenith angle dependence is to be ignored. We have  also successfully  implemented 
the ANN-based energy  reconstruction  algorithm in our  analysis  chain,  by directly using the ANN generated weight-file,  so that the energy of a $\gamma$-ray like  event could be predicted without  using the ANN software package. Further results regarding the resulting reconstructed energy spectra  of the Crab Nebula and Mrk 421 are presented in the next section. 
\section{Performance  evaluation}
\label{11}
The TACTIC telescope with a prototype camera of 81 pixels was made operational in 1997. The telescope   was successful in detecting  intense TeV $\gamma$-ray flaring activity from the BL-Lac object Mkn-501 in the same year.  This  detection  has  an important historical significance in the field of very high energy $\gamma$-ray astronomy  for being the first ever  observation of a TeV $\gamma$-ray source by  5  independent groups [32].
Since 1997, the TACTIC imaging  telescope camera and its data acquisition system  has been   continuously  upgraded.   Observations on potential $\gamma$-ray sources, however, were continued during this interim period  of upgradation  phase whenever it became  possible.  It was during December 2000- March 2001 that  the TACTIC  imaging telescope  in its full configuration of 349 pixels was able to detect $\gamma$-rays from the Crab Nebula and another  BL -Lac object  Mkn-421 at high statistical significances [33]. The   results  of the Crab  Nebula  observations,  
carried out in on/off-source mode (for 41.5 h/30.9 h) between January 19 - February  23, 2001 are shown in Fig.12. 
\begin{figure}[h]
\centering
\includegraphics*[width=1.0\textwidth,angle=0,clip]{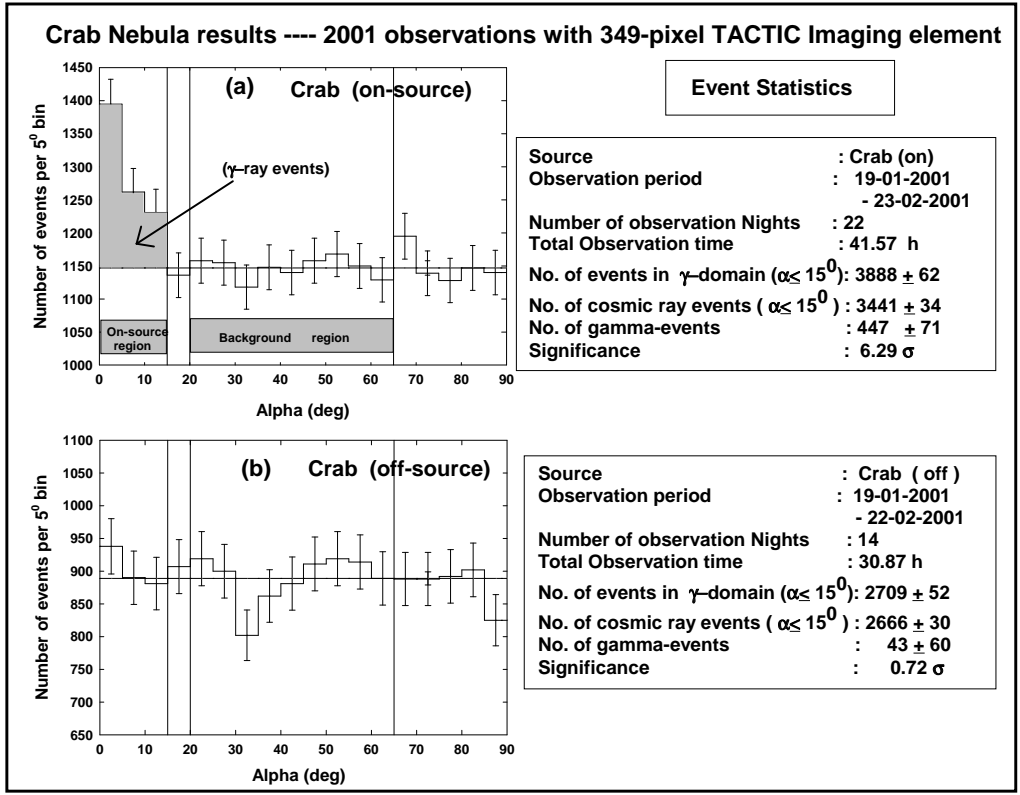}
\caption{\label{fig ---} (a) On-source and (b)  Off-source Alpha-plots of  the Crab Nebula  recorded during  January -February  2001 observation spell  by  349-pixel TACTIC telescope. The number of $\gamma$-ray like events shown in (a) and  indicated by the shaded region are found out to be $\sim$447$\pm$71.}
\end{figure}
The data  was  taken  by using  a  Nearest Neighbour Non-Collinear Triplets topological trigger 
with the innermost 240 pixels (15 x 16 matrix) of the camera   participating in generating the  event-trigger. 
The  3-fold prompt coincidence rate  varied between 3 - 5 Hz. 
Quite reassuringly, a  statistically significant excess of $\sim$6.3$\sigma$  is seen 
in the on-source case (Fig.12a) with respect to the corresponding  background level estimated from 20$^0$ $\le$ $\alpha$ $\le$ 65$^0$ data.  The number of $\gamma$-ray like events in the $\gamma$-domain 
($\alpha$$\le$15$^0$) turns  out to be $\sim$447$\pm$71.  The  corresponding off-source $\alpha$-plot, shown in Fig.12b is  in good agreement with the expected flat distribution.
\par
During the last few  years  regular observations were  taken on a number of potential $\gamma$-ray sources viz., 1ES2344+514, PSR 0355+54, ON 231, H1426, Mrk-421, Mrk-501 etc. The results of these  observations  are presented in [34-36]. While we have so far been  partially successful in  improving the telescope sensitivity   by  fine-tuning some of the critical parameters, like, the coincidence gate width and the trigger field of view,   there  is  still  some  scope to improve telescope performance further   by  making use of better event characterization methodology. 
At present the  telescope has a 5$\sigma$ sensitivity of detecting Crab Nebula in 25 hours of observation time.
This sensitivity  figure  has  been  obtained  by  analyzing   recent  data  on the  Crab Nebula  for $\sim$101.44 h between November 10, 2005 - January 30, 2006   which  has yielded  an excess of $\sim$(839$\pm$89) $\gamma$-ray events with a statistical significance of $\sim$9.64$\sigma$. The data  were  taken  by using  the   Nearest Neighbour Pair topological   trigger 
with the innermost 121 pixels (11 x 11 matrix) of the camera   participating in generating the  event-trigger. 
The  corresponding average $\gamma$-ray rate  for the above  observations  on the Crab Nebula turns out to be $\sim$(8.27$\pm$0.88)h$^{-1}$  at $\gamma$-ray energies of $\geq$ 1.2 TeV.
The  differential energy spectrum of the Crab Nebula  as measured by the TACTIC telescope  is shown  in Fig.13a and  can be  represented by  power law   $(d\Phi/dE=f_0 E^{-\Gamma})$  with  $f_0=(2.74\pm0.19)\times 10^{-11} cm^{-2}s^{-1}TeV^{-1}$  and $\Gamma=2.65\pm0.06$.  Excellent matching of this spectrum with that obtained  by the HEGRA  and Whipple groups [31,37]  reassures  that the  telescope  is functioning  properly. 
\par 
Our extensive  observations of Mrk 421  on 66 clear nights from December 07, 2005 to April, 30, 2006, totalling $\sim$ 202 hours of on-source observations  have also indicated  presence of a TeV $\gamma$-ray flaring activity from the source [38]. The time-averaged differential $\gamma$-ray spectrum in the energy range $1-11$ TeV  is shown in Fig.13b for the data taken between December 27, 2005 to February 07, 2006  when the source was in a relatively higher state as compared to the rest of the observation period. 
\begin{figure}[h]
\centering
\includegraphics*[width=1.0\textwidth,angle=0,clip]{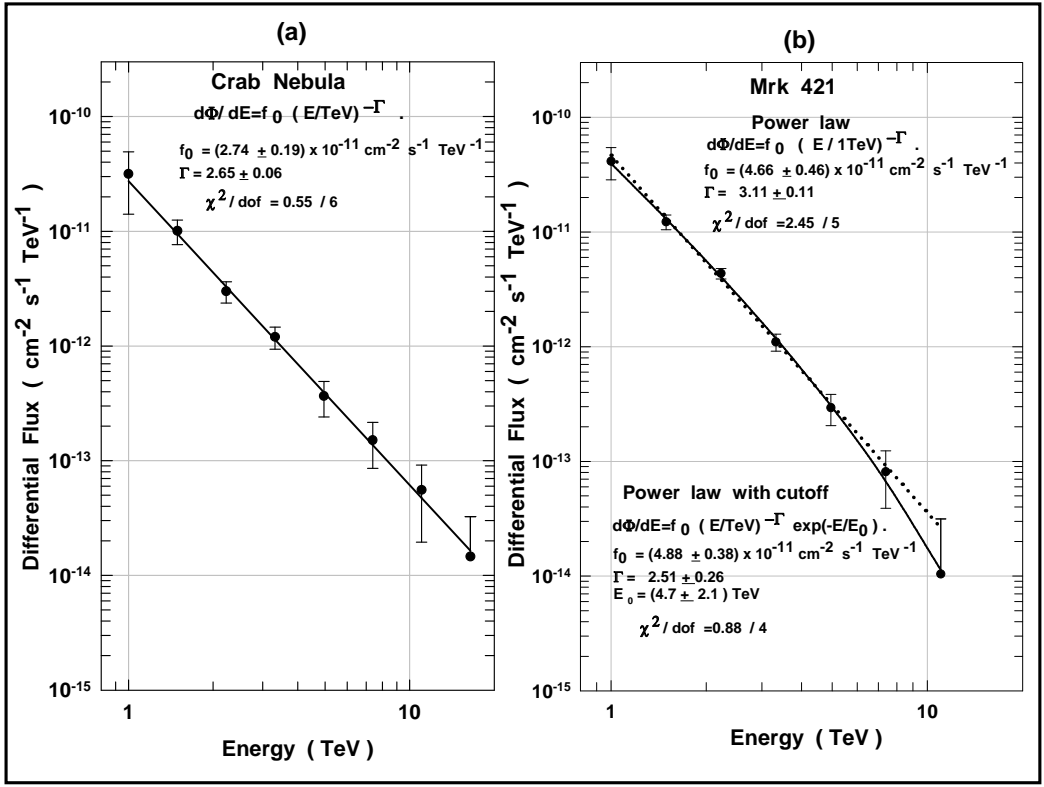}
\caption{\label{fig ---}  (a)The differential  energy spectrum of the Crab Nebula as measured by the TACTIC  telescope.  (b) Differential  energy spectrum of  Mrk 421 for the data collected between 
December 27, 2005 - February  07, 2006 when the source  was in a high state.}
\end{figure}
Analysis of this  data spell, comprising about $\sim$97h reveals the presence of a $\sim 12.0 \sigma$ $\gamma$-ray signal  with  daily flux of $>$ 1 Crab unit  on several days. 
Although not statistically very significant, possible  signature  of  a  exponential cut-off  in the spectrum  at an energy of  $\sim(4.7\pm2.1)TeV$  is also  seen in Fig.13b.  Difficulties like limited $\gamma$-ray event statistics  coupled with rather large  error bars do not  allow us  to claim the cutoff feature at a high confidence level.   
Other  relevant  details  regarding  analysis of the Mrk 421 data,   light curve of the source during our observation period and a novel artificial neural network-based energy estimation procedure  for determining the energy spectrum of a candidate $\gamma$-ray source can  be seen in  [38]. 
\section{Conclusions}
\label{12}
The 349-pixel TACTIC imaging telescope, 
has been  in operation at  Mt.Abu, India  since 2001  and   has  so far 
detected $\gamma$-ray emission from  the Crab Nebula, Mrk 421 and Mrk 501. 
While excellent matching of the Crab Nebula  spectrum with that obtained  by  other  groups  reassures   us  that the  telescope subsystems are functioning  properly,  the inferred sensitivity level  of  1 Crab Unit at $\sim$ 5.0 $\sigma$ in $\sim$25h  needs to be further improved.
Apart from  validating  the stability of the TACTIC  subsystems  directly with $\gamma$-rays from this source,
matching of the  Crab Nebula spectrum also  validates the full analysis chain, including  the inputs used from the Monte Carlo simulations, like,  effective area and $\gamma$-ray acceptance factors and the energy reconstruction procedure.  Furthermore, keeping in view  that the  weak signature of possible cutoff  energy  in the energy spectrum of   Mrk 421, inferred from  our  observations,  is fairly consistent with the  observations of  other groups [39-41], we believe  that there is  considerable scope  for the TACTIC  telescope  to monitor  similar  TeV $\gamma$-ray emission  activity from  other  active  galactic nuclei  on a long term basis.  Participating in  multi-wavelength observation campaigns  on  various active galactic nuclei  and determining  the energy spectrum of these sources (both at low average flux levels of $<$ 1 Crab Unit  and from intense flares of  $>$2 Crab units)  will  be one of the main scientific  objectives  for which  the TACTIC  telescope  will be used  during the next few years of its operation. The data collected on these sources will  also be utilized to understand the  $\gamma$-ray production mechanisms of these objects and  absorption effects at the source or in the intergalactic medium due to interaction of gamma-rays with the extragalactic background photons [42,43]. 
\section{Acknowledgements}
\label{13}
We  would like to dedicate  this paper to the memory of our departed team leader late Dr. C.L.Bhat  who initiated the $\gamma$-ray astronomy programme at Mt. Abu  but  did not live to see it fructify.  Unfortunately, he  met with a fatal road accident while returning from Mt.Abu  on  December 17, 2001. 
The authors would also like to thank colleagues at  the   Centre for Design and Manufacture,  the Electronics  Division and Astrophysical Sciences Division  of our centre  who have contributed during  the various stages of the planning, fabrication, installation and operation of the telescope. We  would  like to thank Dr.V.G.Sinitsyna and other colleagues from  the P.N.Lebedev Institute, Moscow for  many useful discussions and also for providing  us  the design details  of the  SHALON telescope.

\end{document}